# Vehicles, Pedestrians, and E-bikes: a Three-party Game at Right-turn-on-red Crossroads Revealing the Dual and Irrational Role of E-bikes that Risks Traffic Safety

Gangcheng Zhang[a], Yeshuo Shu[a], Keyi Liu[a], Yuxuan Wang[a], Donghang Li[a], Liyan Xu[a*]

a. College of Architecture and Landscape, Peking University, Beijing 100871, China
b. College of Urban and Environmental Sciences, Peking University, Beijing 100871, China

* Corresponding authors

## ABSTRACT

The widespread use of e-bikes has facilitated short-distance travel yet led to confusion and safety problems in road traffic. This study focuses on the dual characteristics of e-bikes in traffic conflicts: they resemble pedestrians when interacting with motor vehicles and behave like motor vehicles when in conflict with pedestrians, which raises the right of way concerns when potential conflicts are at stake. Using the Quantal Response Equilibrium model, this research analyzes the behavioral choice differences of three groups of road users (vehicle-pedestrian, vehicle-e-bike, e-bike-pedestrian) at right-turn-on-red crossroads in right-turning lines and straight-going lines conflict scenarios. The results show that the behavior of e-bikes is more similar to that of motor vehicles than pedestrians overall, and their interactions with either pedestrians or motor vehicles do not establish a reasonable order, increasing the likelihood of confusion and conflict. In contrast, a mutual understanding has developed between motor vehicles and pedestrians, where motor vehicles tend to yield, and pedestrians tend to cross. By clarifying the game theoretical model and introducing the rationality parameter, this study precisely locates the role of e-bikes among road users, which provides a reliable theoretical basis for optimizing traffic regulations.

## 1. INTRODUCTION

E-bikes, characterized by pedal assistance, a maximum speed of 25 km/h, and an electric power output of no more than 250 W, have decreased conventional bicycles' physical constraints and distance limitations (*1*). These bikes offer convenience and agility compared to motor vehicles, so they have gained popularity in East Asia, Southeast Asia, North America, and Europe (*2*). They have become a preferred alternative for both commuting and short-distance travel (*3, 4*). However, researchers are increasingly concerned about the potential road safety issue for e-bikes (*5*), as the increased speed and weight of e-bikes compared to conventional bicycles have increased traffic accident risks (*6, 7*).

In particular, the potential traffic accident risks involving e-bikes are twofold (*3*). On the one hand, riders of e-bikes are at greater risk of injury or death when colliding with motor vehicles. Therefore, they are viewed as a vulnerable group of road users. On the other hand, in China, e-bikes are classified as non-motorized vehicles and must travel in non-motorized lanes (*8*). Although this regulation has lowered the chances of collisions between e-bikes and motorized vehicles, it has also led to more frequent interactions with pedestrians and traditional bicycles, creating new safety challenges. Studies show that in the event of clashes between e-bikes and pedestrians, the latter face higher risks of harm (*9*). Overall, e-bikes display a unique dual role in traffic safety terms: in the event of a collision with a motor vehicle, they are seen as vulnerable road users, much like pedestrians; however, in clashes with pedestrians, they behave like motor vehicles, taking on a dominant role among road users.

The regulatory environment in China has augmented the dilemma. Although the law in China tends to favor and protect vulnerable road users by often prioritizing their right of way, this principle has not been effectively implemented in traffic management practices in certain cases (*8*). China's traffic safety regulations state that as dominant road users, motor vehicles must yield to vulnerable non-motorized vehicles such as e-bikes and bicycles. However, no explicit legal provision exists on whether e-bikes should yield to pedestrians. In reality, e-bikes often travel at varying, sometimes dangerous speeds on roads shared with pedestrians, where effective regulation is lacking (*9, 10*). This void undoubtedly increases safety risks for both parties. Therefore, it is essential to conduct an accurate analysis of the interactions, conflicts, and decision-making processes between vehicles, e-bikes, and pedestrians to optimize existing traffic rules.

To do so, we must focus on one particular situation when all traffic participants, including pedestrians, e-bikes, and vehicles, have roughly the same right of way. Fortunately, the right-turn-on-red (RTOR) crossroads present such a chance. On the RTOR crossroads, right-turning traffic is intertwined with straight-going traffic (*11*). Since all three categories above have the right of way, there is intense competition for space at these locations, resulting in higher accident risks (*12*). Specifically, in this complex traffic scenario, there are three typical interactions between road users: right-turning vehicles with straight-going e-bikes, right-turning vehicles with straight-going pedestrians, and right-turning e-bikes with straight-going pedestrians. These three situations are structurally similar, with the straight-going party generally being more vulnerable than the right-turning one. Specifically, e-bikes are relatively vulnerable in the first interaction but take a stronger position in the third. Therefore, this scenario allows for a concrete manifestation of the boundary-crossing characteristic of e-bikes.

Game theory-based models are a powerful tool to quantitatively analyze the decision-making processes of all traffic participants on such occasions, which have been extensively employed in explaining, modeling, and simulating road user behavior (*13–16*). Previous research has utilized static game theoretical models to provide general insights into road user behavior (*17*), and sequential games to further account for the dynamic decision-making process of road user interactions (*18*). Nevertheless, both static and sequential game theoretical models assume that players possess a high degree of rationality, which can be unrealistic for real-world situations. To remedy the problem, an emerging Quantal Response Equilibrium (QRE) model was introduced to capture uncertainty and errors in the decision-making process. As an extension of the standard Nash equilibrium, QRE combines discrete choice behavior with game theoretical models, relaxing the requirements on players' rationality and allowing for errors in their choices (*19*). Recent research has shown that QRE can accurately model the decision-making behavior of various road users (*20, 21*).

This paper uses a QRE framework to empirically analyze the interaction behavior of the three road users at the RTOR crossroads. The research focuses on three questions:

1. How can a game theoretical model be constructed to describe the three interactions uniformly?
2. Are there significant differences in the game outcomes among the three interactions?
3. Are there noticeable differences in the choices made by e-bikes when playing two different roles?

## 2. METHODOLOGY
### 2.1. Game theoretical model Formulation
#### 2.1.1. Definitions

A standard is needed to determine the beginning and end of a game when dealing with different types of crossroads and players. Based on previous research, the game is considered to begin when both players enter the "influencing area," typically indicated by prominent traffic signs (22). Specifically, a game starts when the right-turning player has entered the right-turn lane area and the straight-going player has entered the curb area or crosswalk. The game ends when both players leave the conflict zone (**Figure 1**). In addition, a game is valid when each player enters the Influence area and the other player has not left or is not too close to the conflict zone.

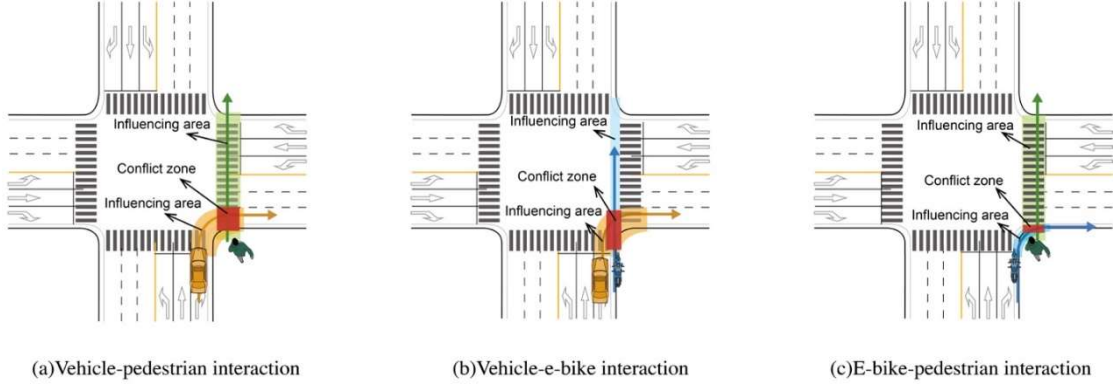

(a)Vehicle-pedestrian interaction  (b)Vehicle-e-bike interaction  (c)E-bike-pedestrian interaction

**Figure 1 Illustrations for the influencing area and conflict zone**

The crossroads interactions are modeled as a static game, where right-turning and straight-going players simultaneously make one-time decisions. Every player has two available strategies. For right-turning e-bikes and vehicles, they have a choice to either "yield" or "not yield," while for straight-going e-bikes and pedestrians, they can either "cross" or "stay." The three interactions share the same model structure, which can be formulated as:

$$G = \langle N, \{S_i\}_{i=1}^n, \{u_i\}_{i=1}^n \rangle \tag{1}$$

Where $N = \{1 = R, 2 = S\}$ is the set of players, $S_i = \{s_{i1}, s_{i2}\}$ is the set of player i's strategies, consisting of two pure strategies. $S_1 = \{1 = yield, 2 = not\ yield\}$ and $S_2 = \{1 = cross, 2 = stay\}$. $u_i: S \to \mathbb{R}$ is the payoff function of player i, where $S = S_1 \times S_2$ is the set of all players' pure strategy profiles. Then, a mixed strategy of player i can be denoted as $\sigma_i = \{\sigma_i(s_{i1}), \sigma_i(s_{i2})\}$, where $\sigma_i(s_{i1}) + \sigma_i(s_{i2}) = 1$, $\sigma_i(s_{i1}) \geq 0$, $\sigma_i(s_{i2}) \geq 0$. It is noteworthy that the player types and beliefs which are introduced in previous studies are redundant (21, 23).

In QRE, individuals are assumed to have disturbances of their utility that other players or researchers cannot observe. These disturbances are used to describe uncertainty in individual behavior (24). At this level, the game becomes an incomplete information game. For any given mixed strategy profile $(\sigma_1, \sigma_2)$, the expected utility of player *i* adopting a pure strategy $s \in S_i$ when the other player continues with $\sigma_{-i}$ is:

$$\widehat{EU}_{i,s} = EU_{i,s} + \frac{1}{\lambda}\epsilon_{i,s} \tag{2}$$

where $\widehat{EU}_{i,s}$ is player *i* supposed expected utility, $EU_{i,s} = \sigma_j(s_{j1})u_i(s, s_{j1}) + \sigma_j(s_{j2})u_i(s, s_{j2})$ is player *i*'s deterministic expected utility, $\epsilon_{i,s_j}$ is the privately observed disturbances. The parameter $\lambda$ serves as a measure of rationality, quantifying how players can make errors in evaluating expected utility. When $\lambda = 0$, players cannot discriminate between strategies, leading them to make random choices. When $\lambda \to \infty$, the QRE equilibrium collapses to the standard Nash equilibrium. Compared to previous studies that did not include the rationality parameter, introducing $\lambda$ can improve the model performance and provide information on rational decision-making.

### 2.1.2. Payoff functions

The accuracy and interpretability of the payoff function, whose composition should reflect the factors that influence players' decisions, are essential to the validity of the game theoretical model. However, previous studies did not explicitly state the criteria for selecting variables or simply included many variables and removed insignificant ones stepwise (*20–23*).

As this study aims to compare the results of the three games, a consistent payoff setting is required. Efficiency and safety are considered the main factors affecting road users' decisions supported by common sense and real-world experience (*25, 26*). For two players $i$ and $j$, the player $i$'s efficiency payoff is represented by $V_i/D_i$, which describes how fast the player can pass the conflict zone. Player $i$'s safety payoff is represented by $V_j/D_j$, which describes how fast the other player approaches the conflict zone. These two revenues are only paid in certain situations: players receive efficiency payoffs only if they choose to cross the road, and they face safety losses only if both players cross. In addition, the safety payoff is supplemented by the number of groups, with higher losses when right-turning players conflict with multiple straight-going players. Although variables such as gender and age of pedestrians are observed and recorded, these control factors are excluded from the model for integration into a unified framework. The payoff matrix is shown in **Table 1**.

**TABLE 1 Payoff Matrix of the Crossroads Game**

| Straight-going player (s) | Payoff | Right-turning player (r) | |
|---|---|---|---|
| | | Yield | Not yield |
| Cross | $U_s$ | $\beta_0 V_s/D_s$ | $\beta_0 V_s/D_s + \beta_1 V_r/D_r + \beta_2 N_{group}$ |
| | $U_r$ | $c_2$ | $\beta_3 V_r/D_r + \beta_4 V_s/D_s$ |
| Stay | $U_s$ | $c_1$ | $c_1$ |
| | $U_r$ | $c_2$ | $\beta_3 V_r/D_r$ |

For a given mixed strategy profile $\sigma_R = (p_{yield}, 1 - p_{yield})$ and $\sigma_S = (p_{cross}, 1 - p_{cross})$, the expected utilities of both players' pure strategies are given as:

$$EU_{cross} = \beta_0 V_s/D_s + (1 - p_{yield}) \times \beta_1 V_r/D_r + (1 - p_{yield}) \times \beta_2 N_{group} \quad (3)$$

$$EU_{stay} = c1 \quad (4)$$

$$EU_{yield} = c_2 \quad (5)$$

$$EU_{not\ yield} = \beta_3 V_r/D_r + p_{cross} \times \beta_4 V_s/D_s \quad (6)$$

Including intercept $c_1$ and $c_2$ will make the model more flexible. The variables used in the formula are listed in **Table 2**.

**TABLE 2 Descriptions of Model Variables**

| Variable | Description |
|---|---|
| Vs | The crossing speed of the straight-going player. (m/s) |
| Vr | The crossing speed of the right-turning player. (m/s) |
| Ds | The distance of the straight-going player from the conflict zone. (m) |
| Dr | The distance of the right-turning player from the conflict zone. (m) |
| Ngroup | Size of the straight-going group. |
| Ycross | The straight-going player's decision. (1=Cross, 0=Not cross) |
| Yyield | The right-turning player's decision. (1=Yield, 0=Not yield) |

### 2.1.3. Logit Quantal Response Equilibrium

In practice, the logit QRE is the most commonly used QRE specification. Logit QRE assumes that the $\epsilon_{ij}$ in **Equation 2** are independently and identically distributed according to the log Weibull distribution (*24*). With some algebraic derivation, the choice probability of pure strategies in the model is equal to:

$$p_{cross} = \frac{exp\left(\lambda \cdot EU_{cross}(p_{yield})\right)}{exp\left(\lambda \cdot EU_{cross}(p_{yield})\right) + exp(\lambda \cdot EU_{stay})} = \varphi(\lambda \cdot \Delta EU_s) \quad (7)$$

$$p_{yield} = \frac{exp\left(\lambda \cdot EU_{yield}(p_{cross})\right)}{exp\left(\lambda \cdot EU_{yield}(p_{cross})\right) + exp(\lambda \cdot EU_{not\ yield})} = \varphi(\lambda \cdot \Delta EU_r) \quad (8)$$

Where $\varphi(x) = \exp(x)/(1 + \exp(x))$ is the cumulative function of the logistic distribution. $\Delta EU_s = EU_{cross} - EU_{stay}$ and $\Delta EU_r = EU_{yield} - EU_{not\ yield}$ are the latent expected utility indices.

Consequently, the probability of strategy selection is transformed into the well-known logit model. Under the given $p_{cross}$ and $p_{yield}$, the parameter values from **Equation 3** to **Equation 6** and the rationality coefficient $\lambda$ can be estimated using the maximum likelihood method. The maximum likelihood function $LL$ of the model is given as:

$$LL_s = \sum_{n=1}^{N}\left(Y_{cross} \cdot \ln(\varphi(\lambda \cdot \Delta EU_s)) + (1 - Y_{cross})\ln(\varphi(-\lambda \cdot \Delta EU_s))\right) \quad (9)$$

$$LL_r = \sum_{n=1}^{N}\left(Y_{yield} \cdot \ln(\varphi(\lambda \cdot \Delta EU_r)) + (1 - Y_{yield})\ln(\varphi(-\lambda \cdot \Delta EU_r))\right) \quad (10)$$

$$LL = LL_s + LL_r \quad (11)$$

Solving the $p_{cross}$ and $p_{yield}$ is a fixed-point problem, where $p_{cross} = f(p_{yield})$ and $p_{yield} = g(p_{cross})$. Recently, an Expectation-Maximization (EM) algorithm has been developed and tested. For more information about this algorithm, readers may refer to previous research (*21, 23*). According to the sensitivity test performed, the initial values do not affect the final convergence, which is consistent with the previous research. Hence, we set the initial values of $p_{cross}$ and $p_{yield}$ to 0.5 each. The iteration is considered to converge, usually within 50 rounds, if $\Delta p_{cross}$ and $\Delta p_{yeild}$ are both less than 0.001.

## 2.2. Data

Three sites were selected in Ningbo, China, including a formal-signalized crossroads, a formal-unsignalized crossroads, and an informal-unsignalized crossroads (as shown in **Table 3**). These sites met the following criteria: (1) Sufficient pedestrian, e-bike, and vehicle traffic volume. (2) The three road users share the same roadway space without dedicated right-turn lanes. (3) Adequate sight distance to allow for individual awareness of potential interactions. Ten hours of video were recorded for each site in October 2020. Peak-hour videos were excluded to avoid the impact of traffic congestion.

**TABLE 3 Details of the Observation Sites**

| Site | Formal | Signalized | Lanes | Samples | Location |
|---|---|---|---|---|---|
| Site1 | Yes | No | Three lines per direction | 306 | 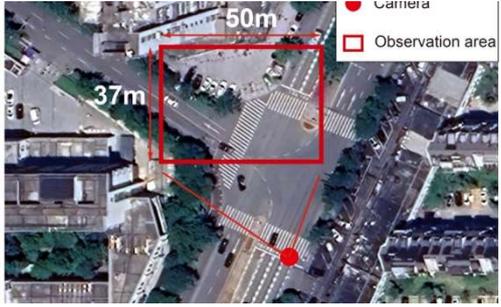 |
| Site2 | Yes | Yes | Three lines per direction | 132 | 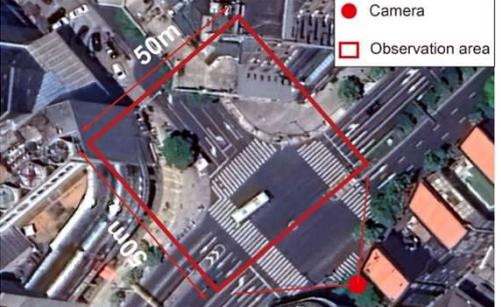 |
| Site3 | No | No | One line per direction | 65 | 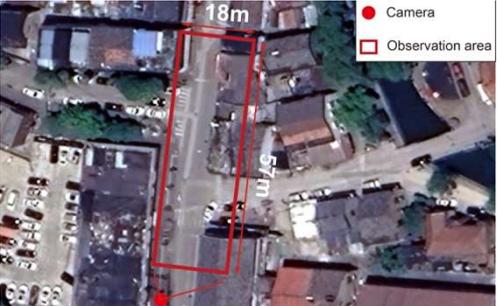 |

The YOLO (you only look once) Version 8.0 (https://github.com/ultralytics) and ByteTrack algorithms (*27*), which are effective methods for multi-target trajectory tracking, were applied for video target recognition and trajectory data extraction (**Figure 2**). As surveillance video images are not orthophotos, a single-adaptive matrix mapping method was used to transform the trajectories from the video image coordinate system to the projective coordinate system (*28*). This approach enabled the precise computation of the kinematic features of road users.

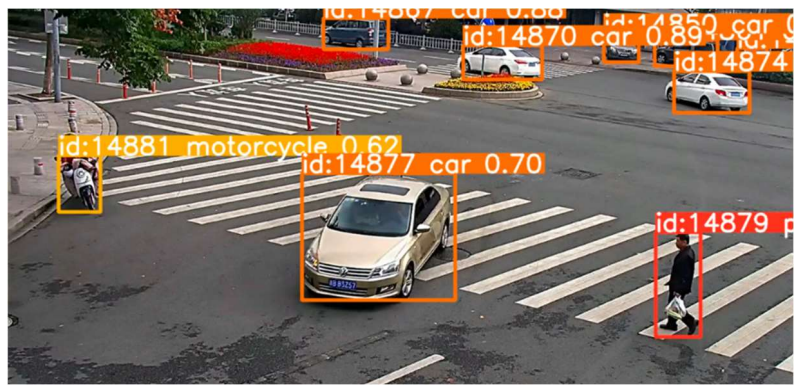

**Figure 2 Trajectory Tracking of Road Users**

Identifying interactions typically requires manual interpretation and judgment. This approach can be costly and potentially lead to biased sample selection. Therefore, a simple algorithm was developed in this paper. Pairs of right-turning and straight-going trajectories were extracted if they intersected spatially and reached the conflict zone within a Post Encroachment Time (PET) threshold of 6 seconds, based on previous research (*22*). Afterward, all extracted trajectory pairs were checked and filtered according to the game definition. We obtained 503 valid interactions, of which 226 were between right-turning vehicles and straight-going e-bikes, 217 were between right-turning vehicles and straight-going pedestrians, and 60 were between right-turning e-bikes and straight-going pedestrians. Notably, the collision volume of e-bikes and pedestrians is small, resulting in a relatively low frequency of their interactions.

## 3. RESULTS
### 3.1. Descriptive Analysis

The descriptive statistics of the model variables and PET are presented in **Table 4**. Several characteristics of the road users can be observed. First, e-bikes had the highest speed whether going straight or turning right, with an average of about 15 km/h. Second, e-bikes and vehicles tended to enter the game from relatively longer distances than pedestrians. These two observations implied that the kinematic features of e-bikes were more similar to those of motor vehicles. Furthermore, fewer e-bikes made yielding decisions when they were the right-turning players. Finally, the PETs of the two interactions involving e-bikes were low, indicating a higher risk of conflict.

**TABLE 4 Descriptive Statistics of Model Variables**

| Variable | Vehicle-pedestrian interaction | | Vehicle-e-bike interaction | | E-bike-pedestrian interaction | |
|---|---|---|---|---|---|---|
| | Mean | Std | Mean | Std | Mean | Std |
| $V_s$ | 1.348 | 0.346 | 4.341 | 1.787 | 1.098 | 0.338 |
| $V_r$ | 3.957 | 1.708 | 3.707 | 1.280 | 4.478 | 1.590 |
| $D_s$ | 6.929 | 2.696 | 14.829 | 6.303 | 3.758 | 2.204 |
| $D_r$ | 19.955 | 7.337 | 14.399 | 4.822 | 9.276 | 5.192 |
| $N_{group}$ | 1.700 | 1.066 | 1.230 | 0.589 | 1.350 | 0.515 |
| $Y_{cross}$ | 0.783 | 0.413 | 0.637 | 0.482 | 0.517 | 0.504 |
| $Y_{yield}$ | 0.650 | 0.478 | 0.465 | 0.500 | 0.333 | 0.475 |
| PET | 3.153 | 1.319 | 2.423 | 1.189 | 2.523 | 1.469 |

**Figure 3** further presents the relationship between the players' speed, distance from the conflict zone, and strategy choices. The "cross" and "not yield" choices tended to cluster in the upper-left region, while the opposite choices tended to cluster in the lower-right area. This observation suggested that players' strategy choices were influenced by the relative magnitudes of their speed and distance, which supported the decision to use the ratio of speed and distance as a variable to describe players' motion states.

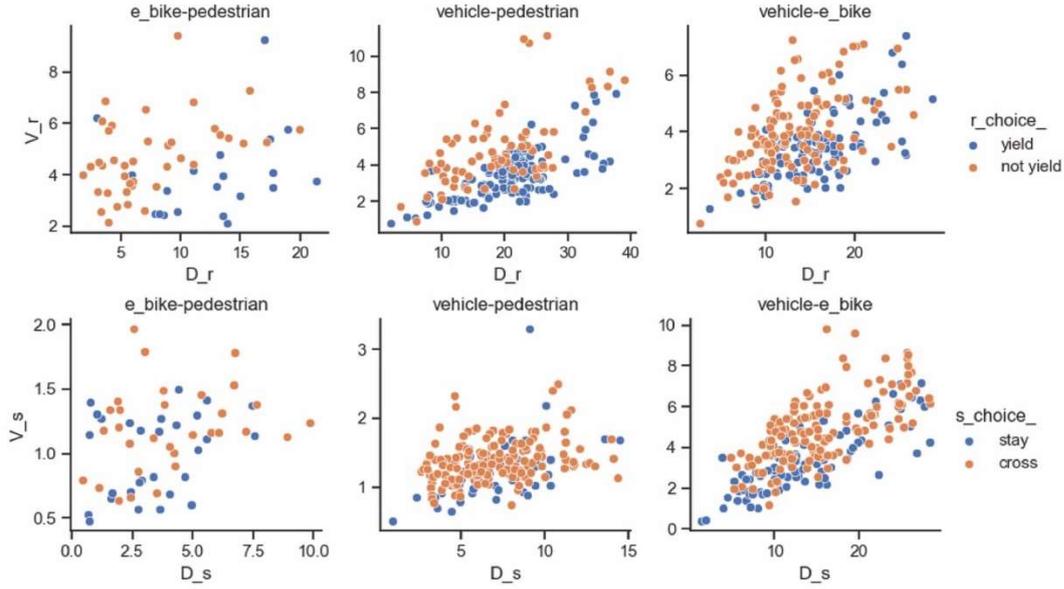

Figure 3 Scatterplots of Distance and Speed by Strategy Choice

### 3.2. Model Estimations and Comparisons

The QRE convergence results of the three games are presented in **Figure 4**. The curves of $p_{not\ yield} = 1 - p_{yield}$ are plotted instead of $p_{yield}$ to highlight the differences between the three games. In the vehicle-pedestrian game, $p_{cross}$ converged to 0.832, while $p_{yield}$ converged to 0.592. In the vehicle-e-bike game, $p_{cross}$ converged to 0.691, while $p_{yield}$ converged to 0.395. In the e-bike-pedestrian game, $p_{cross}$ converged to 0.506, while $p_{yield}$ converged to 0.243. The parameter estimation results are listed in **Table 5**. The bootstrap standard deviations and z-values were used to describe the distribution and significance of the estimated parameters.

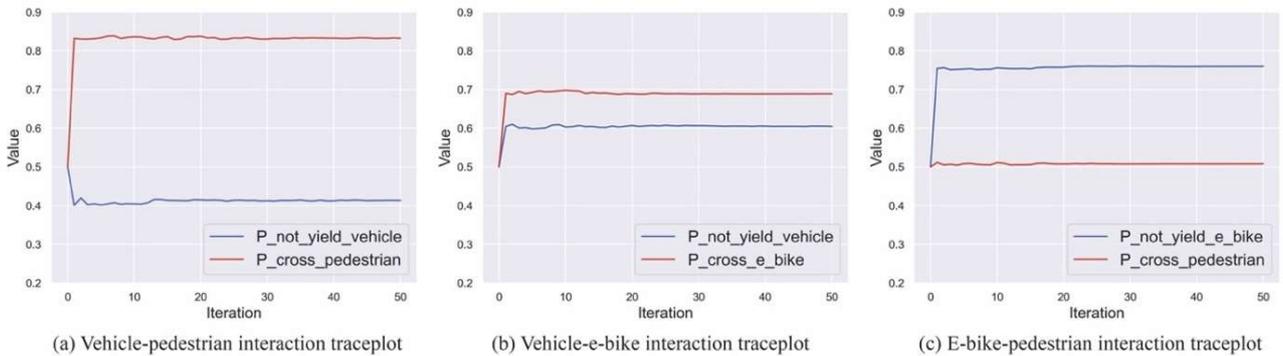

Figure 4 Traceplots of Iterations

**TABLE 5 Estimation Results of the Game Theoretical Model.**

| Variable | Vehicle-pedestrian game | | Vehicle-e-bike game | | E-bike-pedestrian game | |
|---|---|---|---|---|---|---|
| | Coefficient | Bootstrap z | Coefficient | Bootstrap z | Coefficient | Bootstrap z |
| EUcross | | | | | | |
| Vs/Ds | 0.822 | 1.458 | 1.804** | 6.582 | -0.144 | -0.475 |
| Vr/Dr | -2.797** | -3.087 | -1.907** | -5.324 | -0.983* | -2.12 |
| Ngroup | 0.377** | 2.688 | 0.255* | 2.503 | 0.170 | 0.632 |
| Constant | -0.107 | -0.620 | 0.279 | 0.965 | -0.403 | -1.237 |
| EUyield | | | | | | |
| Vs/Ds | -0.849 | -1.568 | -1.937** | -9.277 | -0.321 | -0.508 |
| Vr/Dr | 4.182** | 6.506 | 2.812** | 16.138 | 1.510* | 2.238 |
| Constant | 0.838** | 6.759 | 0.300** | 3.993 | 0.427 | 1.397 |
| Rationality | | | | | | |
| $\lambda$ | 5.342** | 10.899 | 4.445** | 21.976 | 2.177** | 4.154 |

\* significant at the 5% level; \*\* significant at the 1% level

From the parameter estimation results, it can be concluded that the decision-making behavior of e-bikes was more akin to that of vehicles but differed from that of pedestrians. Pedestrians were predominantly concerned about the safety threat posed by right-turning players, while the effect of efficiency on their decision-making was not significant. Moreover, the presence of pedestrians had a limited impact on the safety of right-turning players due to their low speed and collision volume. Conversely, in the vehicle-e-bike game, both players' decisions were significantly influenced by each other's safety threats and their efficiency.

Several pieces of evidence demonstrated that neither vehicles nor pedestrians reached a mutual understanding with e-bikes, reducing overall traffic efficiency and increasing road safety risks:

In the vehicle-pedestrian game, a clear order developed: pedestrians tended to cross, while vehicles tended to yield. However, no orders occurred in the game with e-bikes: Pedestrians' willingness to cross and vehicles' willingness to yield decreased. E-bikes, on the other hand, consistently preferred to cross. Given these situations, consider cases of conflict *(cross, not yield)* and confusion *(not cross, yield)*:

$$p_{conflict} = (1 - p_{yield})p_{cross} \quad (12)$$

$$p_{confusion} = p_{yield}(1 - p_{cross}) \quad (13)$$

The $p_{conflict}$ of the vehicle-pedestrian game is 0.339 and $p_{confusion}$ is 0.099. The $p_{conflict}$ of the vehicle-e-bike game is 0.418 and $p_{confusion}$ is 0.122. The $p_{conflict}$ of the e-bike-pedestrian game is 0.383 and $p_{confusion}$ is 0.120. Thus the game with e-bikes increased the likelihood of conflict and confusion compared to the clean vehicle-pedestrian game.

By comparing the rationality parameters of the three games, it was evident that the e-bike games showed relatively lower rationality than the vehicle-pedestrian game. This result showed that the likelihood of errors, misunderstandings, and irrational decisions increased when vehicles and pedestrians interacted with e-bikes.

## 4. DISCUSSION

The expected social outcome of the road crossing game should be protecting vulnerable road users. However, our findings showed that two games involving e-bikes performed poorly. In the vehicle-e-bike game, both players compete to cross the road, increasing the risk of conflict. In the e-bike-pedestrian game, pedestrians yield to e-bike riders. The root of the problem lies in that e-bikes, although nominally considered vulnerable road users, exhibit relatively aggressive behavior.

Regulating and guiding the behavior of e-bikes is essential to solving this problem. There are two possible measures: First, limiting the speed of e-bikes should be considered. According to our dataset, over 30% of e-bikes had a maximum speed of more than 20 km/h, and around 13% of e-bikes had a maximum speed exceeding 25 km/h. The high velocity puts pressure on other road users and affects their choice of strategy. Second, social norms and traffic regulations should be introduced. From a game theoretical model perspective, these rules can affect the players' payoff and thus influence their decision-making process. Empirical studies have also shown that familiarity with traffic rules can significantly impact the behavior of e-bike riders (*29*).

The individual characteristics of vehicles and e-bikes, especially motor vehicles, are challenging to observe, which leads to the exclusion of pedestrian control variables in the payment functions. This trade-off sacrifices some of the explanatory power. In **Figure 3**, the choice of "cross" for pedestrians does not show a clear concentration in the upper-left corner, which may be explained by pedestrians' preferences.

In addition, players' choices still require manual judgment, which may introduce bias. A possible solution is to develop a more comprehensive game recognition algorithm based on the proposed interactive identification algorithm, which can significantly reduce manual interference. This is an interesting direction for further research but is beyond the scope of our current study.

## 5. CONCLUSION

This research employed QRE to explore behavioral differences in right-turning and straight-going interactions at RTOD crossroads among three interaction groups, including vehicle-pedestrian, vehicle-e-bike, and e-bike-pedestrian. The result shows that:

1. A clear and well-understood order was established for the vehicle-pedestrian game. Vehicles yielded to pedestrians, and pedestrians chose to cross accordingly.

2. However, in the two games involving e-bikes, the order failed to establish. Players in these games did not make the most rational choices, increasing the likelihood of conflict and confusion. Measures should be taken to regulate the behavior of e-bikes.

3. Regarding kinematics and decision making, e-bikes are more similar to vehicles than pedestrians.

This study has several limitations. The number of observation sites and interactions, especially the interactions between e-bikes and pedestrians, can be further increased to obtain more robust results. As more data is added, the model results can be further used to predict and validate road user behavior.

## ACKNOWLEDGMENTS

The authors thank the Bureau of Urban Management and Law Enforcement of Jiangbei District, Ningbo, for their support in securing data accessibility and accommodating necessary research conditions.

## AUTHOR CONTRIBUTIONS

The authors confirm contribution to the paper as follows: study conception and design: Liyan Xu, Gangcheng Zhang, Yeshuo Shu; data collection: Liyan Xu; analysis and interpretation of results: Yeshuo Shu, Gangcheng Zhang, Keyi Liu, Yuxuan Wang, Donghang Li; draft manuscript preparation: Gangcheng Zhang, Yeshuo Shu, Keyi Liu, Yuxuan Wang. All authors reviewed the results and approved the final version of the manuscript.